\documentclass[letterpaper]{JHEP3}

\usepackage{epsfig}
\usepackage{amsmath}

\title{On the existence and dynamics of braneworld black holes}

\author{A. Liam Fitzpatrick, Lisa Randall, Toby Wiseman\\
Jefferson Physical Laboratory, Harvard University \\
Cambridge MA 02138, USA \\
E-mail: \email{fitzpatr@fas.harvard.edu, randall@physics.harvard.edu, twiseman@fas.harvard.edu}
}

\date{August 2006}

\preprint{HUTP-06/A0036}

\abstract{

Based on holographic arguments Tanaka and Emparan et al have claimed
that large localized static black holes do not exist in the one-brane
Randall-Sundrum model.  If such black holes are time-dependent as they
propose, there are potentially significant phenomenological and
theoretical consequences. We revisit the issue, arguing that their
reasoning does not take into account the strongly coupled nature of
the holographic theory.  We claim that static black holes with smooth
metrics should indeed exist in these theories, and give a simple
example. However, although the existence of such solutions is relevant
to exact and numerical solution searches, such static solutions might
be dynamically unstable, again leading to time dependence with
phenomenological consequences. We explore a plausible instability,
suggested by Tanaka, analogous to that of Gregory and Laflamme, but
argue that there is no reliable reason at this point to assume it must
exist.

}

\begin{document}

\section{Introduction and summary}

In the Randall-Sundrum one-brane model (RS2), a five-dimensional
warped spacetime with a single Minkowski brane and brane-localized
matter, linear perturbations of the Minkowski brane and AdS bulk
appear to a brane observer to be those of a four-dimensional gravity
theory up to energies set by the AdS curvature scale. Hence this model
provides a low-energy dimensional reduction for brane observers, even
though the extra dimension is not compact. However, it is not clear
that the RS2 model exactly reproduces four-dimensional gravity at the
nonlinear level. In this regard, the study of black holes is very
interesting.  Tanaka and Emparan et al. \cite{tanaka,
emparan:quantum,emparan:astro} have argued using holography that large
black holes localized on the brane might behave very differently from
four-dimensional ones, decaying classically much faster than the
conventional quantum Hawking evaporation rate. This would allow this
five-dimensional theory to be distinguished from four-dimensional
gravity at low energies, even though linear perturbations about a flat
brane do not distinguish them until energies 
comparable to the AdS scale.

This possibility might seem puzzling from the perspective of a
four-dimensional effective theory, which contains a four-dimensional
graviton mode with the correct four-dimensional gravitational
interactions. However, since the extra dimension is not compact, the
spectrum of Kaluza-Klein (KK) modes is continuous down to zero
energy. While for linear fluctuations of the flat brane and bulk, one
may still construct a four-dimensional effective theory, the validity
of this theory beyond the linear level is unclear. In particular we
will see that for a non-linear background, such as a black hole, the
mass spectrum of the KK modes becomes significantly altered for modes
with mass around the black hole temperature, thereby distinguishing
the effective theory from pure four-dimensional gravity. The nature of
these strong effects will be critical to determining if a static
stable solution exists.

Tanaka and Emparan et al. \cite{tanaka,emparan:quantum} provided
evidence that classical bulk geometries correspond in the holographic
dual to quantum-corrected black holes. 
This dual theory is 4-d gravity coupled to a gauge theory and matter
fields with $N$ colors, taken in the large $N$ 't Hooft limit, with
large 't Hooft coupling \cite{verlinde,gubser,giddings,duffliu,gk,apr}.
They noted that an interesting question arises in the 4-d holographic
theory when one considers Hawking radiation due to the black hole into
gauge theory fields. If radiation from a free field leads to an
evaporation rate $\sim \hbar$, then radiation from the $O(N^2)$ fields
of the holographic dual theory may yield an effect going as $\sim N^2
\hbar$, which would then persist in the large $N$ limit even as $\hbar
\rightarrow 0$. Tanaka and Emparan et al. pointed out such spontaneous
quantum radiation in the 4-d holographic theory would correspond to a
classical 5-d process in the bulk, which would imply a 5-d black hole
localized to the brane should always classically radiate, and hence
cannot be static. The rate of classical ``evaporation'' or decay of
the time-dependent brane black hole could then be used to place
relatively strong bounds on the effective compactification scale, the
AdS curvature length $L$ \cite{emparan:astro}.  Later work supporting these
ideas appeared in \cite{Anderson:2004md,Casadio:2004nz}.

The key subtlety in the argument of Tanaka and
Emparan et al is that the gauge theory is strongly coupled, with large
't Hooft coupling, and hence it is unclear whether simply multiplying
the free field result by $N^2$ is valid when considering Hawking
radiation.  The aim of this paper is to revisit this issue with this
subtlety in mind. 

We find two consistent options for large localized black
holes. The first is that static stable solutions exist. The second,
although we believe a less likely option, is that static but
dynamically unstable solutions exist. The first is obviously in stark
contrast to the arguments of Tanaka and Emparan et al. The second,
while in detail different - particularly the existence of a static
solution, and the interpretation of the instability as being unrelated
to Hawking radiation - the qualitative result is similar, namely from
the classical 5-d point of view, black holes would shrink at a rate
fast enough to be interesting phenomenologically.

We begin by discussing the existence of static black holes in the 5-d
classical theory. We use a concrete counter-example to the claim of
Tanaka and Emparan et al to argue that in the 4-d holographic dual,
Hawking evaporation is not enhanced by a factor of $O(N^2)$, and is
therefore absent in the $\hbar \rightarrow 0$ limit. Hence we expect
static black holes localized to the brane should exist. 
We review the
likely ``pancake'' geometry of these localized black holes and
interpret it in the 4-d holographic dual as a black hole surrounded by
a thermal halo of strongly coupled CFT matter. 

We then discuss the issue of dynamical stability of such static
solutions. We review using entropy arguments a possible instability
suggested by Tanaka that the end of the black hole in the bulk may be
unstable to breaking off. Since we expect the localized black holes to
have a geometry near the brane which is similar to that of a warped
uniform black string, this instability should be similar to the type found by
Gregory and Laflamme for uniform strings
\cite{gregory:instability}. We explore this mechanism using linear theory
to model the possible localized black hole horizon, concluding that
this instability is unlikely to occur. 

We note that there should be no deep mystery regarding the existence
of localized static black holes. In principle their existence is a
problem in partial differential equations. Whereas analytically there
has been little progress in this \cite{EmparanReall,charmousis,Creek:2006je},
numerical methods have been used to solve the full Einstein equations
for localized static objects on the brane
\cite{wiseman,shinkai,kudoh1}\footnote{See also \cite{Casadio:2002uv}.}. 
The most advanced work is that of Kudoh
\cite{kudoh2}, who indeed finds static localized black holes, but with
radii only up to a few AdS lengths. It remains unclear whether large
localized black holes exist. We note however that the numerical
methods used to find them are presumably difficult to implement for
large black holes due to the scale separation between the radius of
the horizon and the AdS scale, which must still be resolved. Hence it
is unsurprising that very large black holes have not been constructed
numerically, and this certainly cannot be taken as evidence against
their existence. Which of our above options is realized (stable or
unstable static black hole solutions) can only be decisively
determined in the non-perturbative gravity theory.  It remains
interesting to check - and we expect confirm - the stability of black
hole solutions by \emph{dynamical} simulation of the 5-d classical
bulk, using similar methods to \cite{choptuik}.

\section{Existence of localized static black holes}

In this section we discuss the validity of the argument of Tanaka and
Emparan et al. that no static solution can exist. We use a concrete
example of a static black hole in the five-dimensional theory where
one can apply their arguments as a counter-example to their claim. We
then argue in the holographic dual theory that Hawking radiation is
not enhanced by a factor of $O(N^2)$ since the number of asymptotic
states that may be radiated is not enhanced by such a factor. Hence we
expect a static solution should exist. However, whereas spontaneous
Hawking emission to asymptotic states does not occur, there may still
be interesting non-spontaneous processes that are seen at the planar
level. These might lead to instabilities of this static solution,
which we will consider in the section that follows. 

\subsection{A counter-example to the claim of no static black holes}
\label{sec:uniform}

We now present a simple static black hole that falls under the
arguments of Tanaka and Emparan et al. Their arguments imply that in
the holographic theory a black hole must be spontaneously emitting
$O(N^2)$ degrees of freedom, resulting both in the lack of a static
solution in the bulk, as one cannot turn off a spontaneous process,
and also a deformation of the brane geometry from the usual 4-d
Schwarzschild due to the backreaction of this radiation. This example
exhibits neither.

The example we consider is the well-known uniform black string. We note that
without an IR brane, the bulk geometry is singular, and therefore
include one to avoid this subtlety \cite{hawking:adsstring}. Since the
metric,
\begin{equation} 
ds^2 = \left( \frac{L}{z} \right)^2 \left( g_{\mu\nu}(x) dx^\mu dx^\nu + dz^2 \right) 
\label{eq:bstring}
\end{equation}
solves the Einstein equations provided $g_{\mu\nu}(x)$ is Ricci
flat, we can take $g_{\mu\nu}(x)$ to be the Schwarzschild solution,
\begin{equation} 
g_{\mu\nu}(x) dx^\mu dx^\nu = - (1 - \frac{R_S}{r}) dt^2 + (1 - \frac{R_S}{r})^{-1} dr^2 + r^2 d\Omega^2 
\end{equation}
with horizon radius $R_S$ to construct a five-dimensional solution. This
solution is the uniformly warped black string. The UV vacuum brane
resides at $z = L$, and we introduce a vacuum IR brane at $z =
z_{IR}$. While we have included an IR brane, classically we may make
$z_{IR}$ as large as we wish. In particular we will take $L << R_S <<
z_{IR}$. In this limit the arguments of Tanaka and Emparan et
al. would be expected to apply. However, clearly from the form of
\eqref{eq:bstring}, the geometry comprises only a non-trivial zero
mode, with no KK modes excited. In particular the UV brane geometry is
\emph{exactly} that of four-dimensional Schwarzschild. Hence from the
4-d dual perspective, the solution is a static Schwarzschild black
hole with no backreaction from the CFT.

However, although the static solution exists, this black hole-CFT
state is in fact unstable due to the presence of the CFT.  For now we
continue to focus on the existence of the static black hole
solution. We will return to the question of stability in the following
section.

\subsection{Reduction of low energy degrees of freedom from strong coupling}

Let us now consider why Tanaka and Emparan et. al.'s calculation of
Hawking radiation fails.  Tanaka and Emparan et
al. \cite{tanaka,emparan:quantum} proposed a deviation from
conventional four-dimensional behavior by considering the
four-dimensional holographic interpretation of braneworld black
holes. The theory dual to the classical 5-d bulk should be classical
4-d gravity coupled to a large $N$ gauge theory with matter in the 't
Hooft planar limit, with 't Hooft coupling $\lambda = N g_{YM}^2$
large, where the gauge theory is conformal in the IR, although the
brane acts as a UV cut-off. We will simply refer to the dual theory as
a CFT, with this UV cut-off implied. These authors provided evidence
that classical bulk geometries correspond in the holographic dual to
quantum-corrected black holes. Just as the observer on the brane in
the bulk picture sees small deviations from 4-d behavior due to
classical 5-d gravity, an observer\ in the 4-d holographic theory sees
the same deviations from classical 4-d behavior, now due to the
quantum corrections from planar contributions of the gauge theory.

The 4-d quantum corrections in the holographic dual could survive when
we take $\hbar \rightarrow 0$ (note that $\hbar$ is the same for both
theories) if quantum effects coherent amongst the $O(N^2)$ color
degrees of freedom amount to a total effect going as $\sim N^2 \hbar$.
This should be generally true, and can be seen explicitly for the
$\mathcal{N}=4$ case.

Recall that in the large $N$ limit, the 't Hooft coupling
\begin{equation}
g_s N = \left( \frac{L}{l_s} \right)^4 = \lambda
\end{equation}
remains fixed, where $g_s$ and $l_s$ are the bulk string coupling and
length.  $N$ is related to the 4-d Planck length by $N = L/l_4$, so
that in this limit $l_4 = g_s l_s^4/L^3 \rightarrow 0$.
Since $\hbar = l_4^2 / G_4$, with $G_4$ the 4-d Newton constant, the
combined quantity
\begin{equation}
N^2 \hbar = \left( \frac{L}{l_4} \right)^2 \left( \frac{l_4^2}{G_4} \right)
   = \frac{L^2}{G_4}
\end{equation} 
remains fixed
while $\hbar \rightarrow 0$, as we keep $L, G_4$ fixed in RS2.

Consider a 4-d black hole evaporating, with initial radius
$R_S$. Assume for the moment that as Tanaka and Emparan et al claim,
the power emitted by Hawking quanta is $dM/dt \sim N^2 \hbar/R_S^2$,
ie. $N^2$ times the free field result. Then, the evaporation time $T$
remains finite in this $\hbar \rightarrow 0$ limit:
\begin{equation}
1/T \equiv \frac{1}{R_S} \frac{d R_S}{dt}
   = \frac{1}{R_S} \frac{G_4 dM}{dt} = \frac{L^2}{R_S^3}
\end{equation}
We emphasize that this argument
holds provided the Hawking radiation rate does indeed go as $N^2$
times the usual free field result. We now argue that is not the case.

The problem with the argument is that in the holographic dual theory,
there do not necessarily exist $O(N^2)$ dynamical asymptotic degrees
of freedom accessible to radiation from a localized finite temperature
object. This reduction in accessible degrees of freedom derives from
the large 't Hooft coupling when the theory is dual to
gravity.\footnote{Our argument is related, but different to that in
\cite{plasmaballs} which explained the radiation rate of plasma-balls
in large $N$ confining gauge theories.}

We now briefly review the dynamical degrees of freedom for the usual
case of AdS-CFT, where the field theory is $\mathcal{N} = 4$ super
Yang-Mills, and in particular, how the number of colors $N$ manifests
itself in the closed string gravitational dual. Note that although we
are using the results for this particular conformal theory, the
results should be general whenever the closed string dual has a gravity
limit. For details of the review below, the reader is referred to
\cite{Maldacena,Witten1,Gubser,Witten2,Aharony,strassler,sundborg,minwalla:hagedorn}
and references therein.

The situation is best understood for the $\mathcal{N} = 4$ super
Yang-Mills theory on a sphere. Let us take the sphere radius to be
$R_{sph}$. The states in the field theory fall into weakly interacting
states with energies $E R_{sph} << N^2$, and strongly interacting states
with energies $E R_{sph} >> N^2$.

The weakly interacting states are built from a free particle Fock
basis of traces of products of local adjoint fields and their
derivatives. These behave as free particles as in the large $N$ limit
these trace operators commute provided the total number of local
fields in the product is $<< O(N^2)$, implying an energy $E R_{sph} <<
N^2$. These states are best thought of as glueballs, since they
represent excitations of the low temperature confining vacuum for the
theory on a sphere.

The strongly interacting states arise when the number of local fields
in an operator becomes $O(N^2)$, for example considering products of
long trace operators or determinants, and hence its energy becomes $E
R_{sph} \sim O(N^2)$. They can no longer be thought of as a set of weakly
interacting particles since the commutation with other operators is
lost. These states have large energies, but due to the length of the
operators, their density of states is very high, going as
$e^{O(N^2)}$. We term these plasma states, since they describe the
plasma of the high temperature deconfined phases of the theory on the
sphere.

At large 't Hooft coupling the theory is dual to closed string theory
in asymptotic AdS in the gravity limit, $\lambda^{1/4} = L/l_s
\rightarrow \infty$. The glueball states should then correspond to
perturbative string excitations about the vacuum target
spacetime. While in the free theory the separation of energies of the
glueball states is $\sim 1/R_{sph}$, at large 't Hooft coupling the
energy separation scales as $\sim \lambda^{1/4}/R_{sph}$ since the
spacing of the closed string dual spectrum is $\sim 1/l_s$ rather than
$\sim 1/L$. Hence as we approach the gravity limit the entire glueball
spectrum is lifted to infinite energy, apart from the glueballs dual
to the supergravity modes of the string. Then the gravitational
perturbations are dual to only $O(1)$ of the $O(N^2)$ free particle
basis states. In particular the graviton is dual to the field theory
stress tensor. We cannot see `N' for perturbative fluctuations in the
gravity limit, the other $O(N^2)$ perturbative states only becoming
visible when we look at string scale physics.

In our situation we do not restrict the field theory to be
$\mathcal{N} = 4$ super Yang-Mills, the case discussed above. However, in other
field theories known to be dual to string theories the case is
qualitatively similar \cite{strassler}. Since in these cases the
reduction in light glueball degrees of freedom is exactly dual to the
decoupling of string oscillator modes in the string dual when
truncating to the gravity limit, it seems reasonable that this will
occur whenever the string dual to the field theory has such a gravity
truncation.

At large 't Hooft coupling the plasma states correspond to the
non-perturbative black hole excitations in the closed string dual, the
small and large AdS Schwarzschild black holes with radii small or
large compared to the AdS scale $L$. The energies of these black holes
are $E R_{sph} \sim O(N^2)$ translated to the field theory. The large
number density of the plasma states allows them to account for the
$O(N^2)$ entropy of these black holes. At finite temperature, whilst
at large $N$ these states are very massive, their enormous number
density, $e^{O(N^2)}$, may allow them to overcome their exponential
Boltzmann suppression. This occurs for the large black holes which are
rapidly semiclassically spontaneously nucleated in hot AdS above a
critical temperature in the field theory going as $\sim 1/R_{sph}$,
and is reflected in the gravity dual by the negative free energy of
the large black holes. The small black holes, similar to asymptotic
flat space black holes have positive free energy, and hence are
composed of states that are not numerous enough to spontaneously
overcome their Boltzmann suppression.

Having briefly reviewed how field theory states correspond to the dual
gravity physics, and in particular the role that the number of colors
$N$ plays, we now consider radiation from a localized finite
temperature object coupled to the CFT at strong coupling in asymptotic
flat space, such as for our 4-d black hole localized to the brane. We
again are interested in the $\lambda \rightarrow \infty$ limit to
ensure the gravity dual description, and may still consider the theory
on the sphere although we must take the sphere radius $R_{sph}$ to be much
larger than the size of our localized thermal source, the black hole.

Firstly consider the glueball excitations. The $O(1)$ free particle
states dual to the graviton perturbation modes, which have energies
$\sim 1/R_{sph}$, can certainly be radiated. However as described
above the remainder of the $O(N^2)$ glueball states cannot, due to
their enormous energy $\sim \lambda^{1/4}/R_{sph}$, and hence enormous
Boltzmann suppression. In the gravity dual, this corresponds to our
localized black hole thermally radiating gravitons, but not
effectively radiating string oscillator modes due to their enormous
mass.

Secondly we must consider the plasma states. The only way these can be
emitted spontaneously is if they have a sufficiently large degeneracy
to overcome their large $O(N^2)$ energy, and hence large Boltzmann
suppression. In principle this might allow an emission rate that could
account for a classical process in the dual gravity. 

Whilst a thermal emission of such massive objects naively seems
unlikely (see \cite{sorkin} for a discussion of Hawking radiation of
`macroscopic' objects), and our previous counter-example of the warped
string demonstrates explicitly that this does not occur, we now
attempt to give a rough argument why this is so.

We expect that the plasma states dual to small black holes cannot be
spontaneously nucleated by our localized black hole, since they cannot
even be nucleated when the entire theory is put at finite
temperature. So we only consider the possible nucleation of plasma
states dual to large black holes with radius of order $L$ or greater.

Consider now in the bulk the large black holes that might be
`classically emitted' from the localized brane black hole or black
string of the counter-example. Since we are interested in the Poincare
slicing of AdS, there are no finite size static black holes (not
attached to the brane). The only static black holes are infinite in
extent in the brane directions, and hence have infinite energy, and
correspond to a horizon at finite radial position in the
bulk. Obviously such infinite energy objects could not be radiated by
a finite size hot source such as our localized black hole or string.

However, we must also be concerned with black holes of large but
finite radius that are classically emitted near the brane black hole
horizon, and then subsequently fall away from the brane. Such black
holes have a temperature measured on the brane given by their local
horizon temperature, redshifted by the warp factor due to their
distance from the brane. Hence the temperature of a black hole
decreases as it falls away from the brane, which is dual to the
temperature of the thermal plasma decreasing as it expands under its
internal pressure.

Static large black holes in global AdS have a temperature that
increases with their energy. There is a minimum black hole
temperature, attained by a black hole of approximately $L$ in
radius. Such a black hole of radius $L$, within a few AdS lengths of
the brane will then have a temperature measured on the brane given as
$\sim 1/L$. Taking this minimum temperature large black hole and
treating it as a probe in the AdS metric written as
\eqref{eq:bstring}, we can estimate its temperature reduced by the
redshifting at its coordinate distance $z$ from the brane as $T \sim
1/z$.

The important point is that the brane black holes (localized or
string) have lower brane temperature the larger their radius $R_S$,
going as $T \sim 1/R_S$ for $R_S >> L$. Since a thermal object cannot emit
objects hotter than itself, a large 4-d black hole evidently cannot
emit plasma states near its horizon dual to this minimum temperature
black hole within the region $z < R_S$ near the brane. This remains true
for plasma states dual to even larger black holes, which have even
larger temperatures. Now we should consider the plasma states dual to
the minimum temperature large black hole far from the brane. For $z >
R_S$ the brane temperature of these black holes becomes small enough
that the dual plasma states might be emitted. However, for the string
the warp factor means the local horizon radius of the string far from
the brane will be much smaller than $L$, and hence there is little
overlap of the string and the large black hole of radius $\sim L$ to
be emitted, and hence such a process would be suppressed. For the
localized brane black hole, as we shall see later the horizon doesn't
extend further than $z \sim R_S$ into the bulk, and hence again there is
a lack of overlap which would disallow the process. We therefore
conclude that the only plasma states that have any chance to be
emitted are those that are dual to black holes with radius $\sim L$,
at a position $z \sim R_S$ in the bulk. We have no argument to rule
these out, but our counter-example implies that the effects ruling out
the emission of both the $z < R_S$ and $z > R_S$ black holes are still
sufficiently effective at $z \sim R_S$ to stop emission there too.

Since we cannot solve the CFT at large 't Hooft coupling the above
arguments are necessarily heuristic. However, they do show how the
simple thinking that one computes the Hawking radiation rate by
multiplying the free field result by $O(N^2)$ breaks down in the
strongly coupled field theory dual to gravity. We therefore conclude
there is no convincing holographic argument obstructing the existence
of 5-d static black holes localized on the brane due to spontaneous
radiation in the field theory. Furthermore we do not expect any
backreaction from this field theory Hawking radiation to be seen in
the classical gravity dual either. This is in perfect accord with the
static warped uniform string example given above where neither
classical radiation or its backreaction is seen.  However the
situation is interesting when one considers dynamical stability of
static solutions. In the following section we will consider possible
classical instabilities and why instabilities, but not spontaneous
Hawking radiation, might survive in the large $N$ planar limit.

Notice that we have done the analysis in the AdS background that is
relevant to phenomenological applications. It is interesting to note
that whereas the localized black hole appears to exist without
spontaneously radiating in a cold vacuum, if we heat the theory up to
temperatures of order the localized black hole temperature or higher,
this metastability is likely to be effected. In this case, the
holographic picture indicates that the localized black hole should
radiate strongly, at a rate $O(N^2)$, since the 
relevant degrees of freedom at this high temperature are deconfined
with all $O(N^2)$ gluons contributing \cite{Witten2},
so it may evaporate in line with the original arguments of Tanaka and
Emparan et al. This radiation would be due to the emission of the
non-perturbative plasma states of the thermal bath, whose high
temperature `evaporates' the localized hot object.  From the 5-d
perspective, we note that the gravity dual description now includes an
IR horizon in the vicinity of the brane that likely disallows the
static localized 5-d horizon on the brane.
In this case, the brane black hole horizon ends on the bulk black hole and thus, 
instead of rounding off at the tip, looks string-like everywhere.  This 
eliminates the need for the
CFT modes dual to the ``rounding-off'' behavior, since the black string is 
described purely by gravity in the CFT.
Of course, putting the 4-d theory in such a high temperature thermal
plasma bath is interesting but not a case relevant for phenomenology
and we will not discuss this further here.

\subsection{Geometry of localized black holes}

As we now know of no argument, holographic or otherwise\footnote{The
work of Bruni et al \cite{bruni} showed a collapsing shell of matter
which is spherically symmetric on the brane cannot have a static
Schwarzschild exterior unlike in conventional 4-d gravity. This is
often cited as supporting the conjecture that no static localized
black hole exists. Since Birkhoff's theorem no longer applies outside
a spherical brane matter source, due to the presence of the KK modes,
it is entirely to be expected that the exterior brane metric is not
static, while the spherically symmetric KK modes get radiated away. Of
course the geometry will settle down to the static localized black
hole eventually, in no contradiction with the results of
\cite{bruni}.}, that a static localized black hole should not exist,
we review what form they are expected to take. Following Giddings et
al \cite{giddings}, we may estimate the shape of a black hole from the
linear equation governing the field $\phi=1+(z^2/L^2) g_{00}$, i.e. the AdS scalar
Laplacian. We estimate the intrinsic horizon spatial geometry as that
induced on the isosurface where $\phi = 1$. Of
course it is unclear how accurate the linear approximation will be to
the full non-linear solutions, but it is reasonable to expect
qualitative agreement.

\begin{figure}
\centerline{\epsfig{file=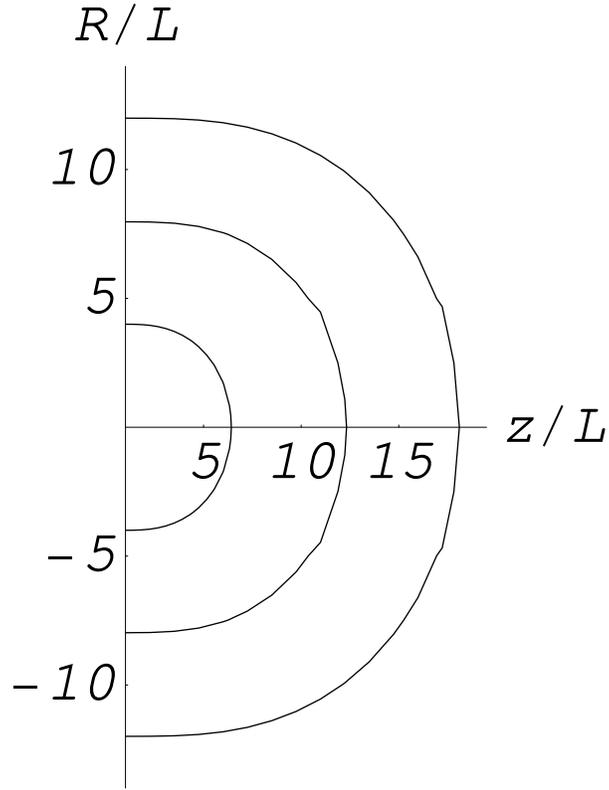,width=3.2in}}
\caption{ Surfaces in AdS in $r,z$ coordinates, computed from linear
theory, whose intrinsic geometry approximates the horizon geometry of
black holes of 3 sizes:  $4L,8L,$ and $12L$. We see these large
localized black holes have horizon geometries simply related by a
global scaling, and extend a coordinate distance $\Delta z \sim 2 R_S$
into the bulk.
\label{fig:iso5}
}
\end{figure}

Taking the AdS coordinates,
\begin{equation}
ds^2 = \frac{L^2}{z^2} \left( - dt^2 + dr^2 + r^2 d \Omega^2 + dz^2 \right)
\end{equation}
the AdS Laplacian is homogeneous in $r$ and $z$,
\begin{equation}
\Box \phi = \left(\frac{z}{L} \right)^2\left( \partial^2_r + \frac{2}{r}  \partial_r + \partial^2_z - \frac{3}{z}  \partial_z \right) \phi
\end{equation}
Taking the brane to be at $z = L$, we must solve the Laplace equation
for Neumann boundary conditions at the brane, but with a static delta
function source at $r = 0$. Following Giddings et al, one then
constructs the metric perturbation from 
$\phi$. The strength of the source determines the size of the black
holes, and hence the position of the locus $\phi = 1$. The brane and
delta function position being at $z = L$ breaks the homogeneous
scaling symmetry of the equation and other boundary conditions, $r,z
\rightarrow \lambda r, \lambda z$ under a change in strength
$\lambda^2$ of the delta function source. However, far from the brane
compared to the AdS scale, the solution does regain this scaling
symmetry. Hence for large black holes, so that the majority of the
isosurface $\phi=1$ is many AdS lengths from the brane, the horizon
isosurfaces have the same shape in the $r,z$ plane, up to the global
scaling of $r$ and $z$. This is illustrated in figure \ref{fig:iso5}
where we plot isosurfaces for a variety of black hole sizes and we see
the larger black holes all have the same shape. The horizon isosurface
extends a coordinate distance approximately $\Delta z \sim 2 R_S$ into
the bulk.

The shape of the isosurface in the $r,z$ plane implies that for large
black holes with horizon radius $R_S >> L$ on the brane, the horizon
geometry near the brane will be approximately that of a warped uniform
string extending into the bulk. Around a proper distance $\simeq L
\log{ R_S / L }$ this warped uniform string ends, being capped off in an
additional proper distance $\sim L$ by a horizon with characteristic
curvature radius $\sim L$. Since this capping off necessarily involves
nonzero mass Kaluza-Klein modes, in the 4-d holographic theory, the
black hole horizon will be surrounded by a strongly coupled halo of
gauge theory matter bound to it. We expect that only glueballs dual to
gravitons may be spontaneously radiated from its surface in the planar
limit. The interesting question is then whether this black hole and
halo state is dynamically stable.

The possible existence of black hole solutions with bound CFT matter
might initially seem surprising. Nonetheless, we know that such bound
states exist in other situations, the simplest example being localized
matter on the brane, such as stars and planets. Similarly, extremal
black holes should take this form. Such solutions perturb the bulk
geometry by sourcing Kaluza-Klein modes, and therefore have a bound
state CFT component from a holographic perspective.

\section{Stability of localized static black holes}

We have shown a concrete example of a static black hole in a 4-d
theory of gravity plus CFT where the Tanaka and Emparan et al
arguments would predict none could exist. However in this theory, as
we discuss below, this exactly Schwarzschild black hole plus trivial
CFT state is actually unstable, and thus may have interesting dynamics
in the spirit of their arguments.  Having concluded that static black
holes exist, we now discuss their dynamical stability.

There are two possibilities. The first is that there is a consistent
stable black hole bound state with the CFT.  The second is that a
solution exists but is unstable. A third possibility, that no static localized
black hole solution exists, is not ruled out by our counter-example
to the arguments of Emparan et al., though it does remove the argument
against their existence.  However, we re-emphasize that static black hole
solutions up to a few AdS lengths have been numerically constructed, so that
static (though perhaps unstable) solutions should exist.
Interestingly, if an instability
were present for the localized black hole, the end result of such a
process may be rather similar to Tanaka and Emparan et al's picture,
namely a rapid loss of energy to infinity, analogous to the result of
the Gregory-Laflamme instability.  We discuss the form such an
instability might take for the brane black hole shortly. We now
briefly discuss the time-dependent evolution that would occur in the
presence of such an instability.

Emparan et. al. suggested that the bulk interpretation of black hole
decay would be classical gravitational radiation {\it near the brane},
through which the black hole slides off the brane into the
bulk. However, this interpretation is clearly problematic in that the
only light mode localized near the brane is the zero mode, and that
mode (in RS2) is not a CFT bound state, but a fundamental normalizable
mode that exists in the presence of the brane. Since any potential
instability should be a consequence of the CFT dynamics, the bulk
holographic interpretation must lie elsewhere.

Tanaka \cite{tanaka} made a different suggestion which is more likely.  The
black hole could decay classically through emission of
higher-dimensional black holes at the tip, where the curvature is
large - of size set by the AdS scale, the size of the tip
region. Specifically, the tip of the black hole, where it extends
farthest away from the brane, is unstable to breaking off.  The
remaining brane black hole would become slightly smaller, and the tip
would become a small higher-dimensional black hole which would then
decay or falls through the horizon of the Poincar\'e patch.  From the
CFT point of view, this would correspond to an instability through
which the black hole could decay much more rapidly than implied by
Hawking radiation.

Tanaka estimated the rate at which the brane black hole would lose
energy to the small black droplets.  The total rate of energy released
is the mass of the black droplets multiplied by the rate of droplet
production.  Although we cannot give a precise rate, from the
perspective of the local 5-d geometry, there is only one scale in the
problem, $L$.  The only other scale, $R_S$ the horizon radius, does not
appear locally.  It seems reasonable to assume that the rate of black
hole production is $L^{-1}$.  The total rate of energy production is
then
\begin{equation}
\frac{dE_{tot}}{dt} \sim \big( M_{drop} \big) \frac{dN}{dt} = (L^2
    M_5^3) (L^{-1})
\end{equation}
with $M_5$ the 5-d Planck mass. An observer on the brane will see this
value redshifted by the factor $(L/R_S)^2$, so the observed evaporation
rate will be
\begin{equation}
\frac{dE_{obs}}{dt} \sim \frac{L^3 M_5^3}{R_S^2} = N^2 A T^4
\end{equation}
where we have used the holographic relation $N^2 = M_5^3 L^3$.  We
have written the last expression in terms of the 4-d parameters, the
area $A$ of the 4-d black hole and its temperature $T=1/R_S$.  
This gives parametrically the same rate of energy loss in the CFT as
Tanaka and Emparan et al's proposal of spontaneous thermal emission of
$O(N^2)$ degrees of freedom. However we note that in light of the
arguments in the previous section, this is not a spontaneous process, 
but rather an instability, and hence can be turned off by fine tuning.

\subsection{Gregory-Laflamme instability}

We now consider the warped black string instability.  Afterward we
will consider a possible analogous instability for the black hole.
The known instability (for the black string) is the Gregory-Laflamme
(GL) instability \cite{gregory:instability,gregory:ads,gregory:entropy}, 
which we now
review. Consider disturbing the warped uniform string metric
\eqref{eq:bstring} by a 4-d tensor perturbation,
\begin{equation} 
ds^2 \rightarrow ds^2 +  h_{\mu\nu}(t, r, z)
dx^\mu dx^\nu .
\label{eq:GLpert}
\end{equation}
Gregory showed \cite{gregory:ads} that the usual vacuum GL instability
of uniform strings generalizes simply to the warped case. This was not
at all obvious as the warped string background is not translationally
invariant. Writing,
\begin{equation}
h_{\mu\nu}(t, r, z) = \chi_{\mu\nu}(t,r) f(z)
\label{eq:GL}
\end{equation}
we take the 4-d tensor $\chi_{\mu\nu}$ to be transverse and traceless
with respect to the 4-d metric $g_{\mu\nu}(x)$. Taking $f(z)$ to be an
eigenmode of the operator, $\partial^2_z - \frac{3}{z} \partial_z$,
with eigenvalue $k^2$, so that,
\begin{equation}
f(z) = \mathcal{A} J_2\left( k z \right) + \mathcal{B} N_2\left( k z \right)
\label{eq:GLprofile}
\end{equation}
and we must choose coefficients $\mathcal{A},\mathcal{B}$ to satisfy
the appropriate boundary conditions. Then the 4-d tensor perturbation
$\chi_{\mu\nu}$ satisfies,
\begin{equation}
\left( \triangle_L^{(4)} + k^2 \right) \chi_{\mu\nu} = 0
\end{equation}
where $\triangle_L^{(4)}$ is the Lichnerowicz operator of the 4-d
metric $g_{\mu\nu}$ given by,
\begin{equation}
\triangle_L^{(4)} \chi_{\mu\nu} = \nabla^2_{(4,S)} \chi_{\mu\nu} + 2 R^{(4)~\alpha~\beta}_{~~\mu~\nu} \chi_{\alpha\beta} = 0
\label{eq:lich}
\end{equation}
with $\nabla^2_{(4,S)}$ the 4-d metric scalar Laplacian, and
$R^{(4)}_{\alpha\beta\mu\nu}$ the 4-d metric curvature, with indices
raised and lowered with respect to the 4-d metric $g_{\mu\nu}$.

This is exactly the equation one obtains for a non-warped
string. Hence as in that case, for $k < k_c = 0.45/R_S$ one
finds modes with an exponentially growing $t$ dependence, leading to the
familiar horizon instability. 
 If we consider an IR brane, then the spectrum of
allowed $\lambda$ becomes quantized, although provided that $z_{IR} >>
R_S$, these include unstable modes.

\subsection{Holographic interpretation of instabilities}

We see from the warped uniform string example  that the
CFT state may be dynamically unstable. There is no contradiction with
our statements about Hawking radiation, since this instability is not a spontaneous
radiative process, but rather is simply a result of having an
energetic instability. The GL instability of gravity implies the
existence of certain tachyonic modes in the gauge theory
description. In the gravity these modes are perturbative graviton
states, and hence in the gauge theory correspond to tachyonic
glueballs. Note that these tachyonic glueball states are quite specific
in form, being spherically symmetric, and exponentially localized near
the horizon. This instability is not spontaneous, as by suitable fine
tuning the system can be prepared to stay static for as long as we
wish. However, when perturbed the occupation number in these tachyonic
glueball modes will grow simply because it is energetically
favorable for this to happen. This will continue until a large
collective behavior is produced.
Thus we conclude that dynamical instabilities of the CFT vacuum may
lead to interesting dynamics in the planar limit, but spontaneous
radiation is not seen in this limit.  The Emparan et al argument
fails for spontaneous emission of radiation, but interesting
effects analogous to their original claims might be possible for
non-spontaneous energetic reasons.

\subsection{A possible instability}

So let us now consider the possibility of a GL-like instability for
the localized brane black hole in the bulk.  The majority of the
proper distance of a large localized black hole with brane radius $R_S
>> L$ appears like the warped uniform string, extending roughly $~ L
\log{R_S/L}$ into the bulk, and it is only in the last $L$ proper
distance that the geometry deviates from the uniform string and caps
off.

The GL instability is the only dynamical instability known for static
black holes, and certainly would account for the decay of the black
hole that was considered above. Gregory and Laflamme argued that for a
uniform string the natural end state of the instability in vacuum is
an array of black holes. This remains an issue of controversy due to
the result of Horowitz and Maeda \cite{horowitz}. However, following
numerical simulation of the instability by Choptuik et al
\cite{choptuik}, it has recently been argued that Gregory and
Laflamme's original picture is likely to be correct
\cite{marolf,garfinkle,kol:caged}. If the uniform string region of a localized
black hole were long enough, one would expect a similar instability to
exist on it, whose dynamics would result in the uniform neck breaking
up, and the segments not connected to the brane falling into the bulk.

In the CFT such an instability would manifest itself through the existence of tachyonic
glueball states. Perturbing the static black hole and its halo would
result in a condensation into these tachyonic states. The result would
be violent emission of glueballs (corresponding to gravity waves in
the 5-d gravity), and expanding cooling shells of thermal gluon plasma
(corresponding to the small black holes falling away from the brane).

\begin{figure}
\centerline{\epsfig{file=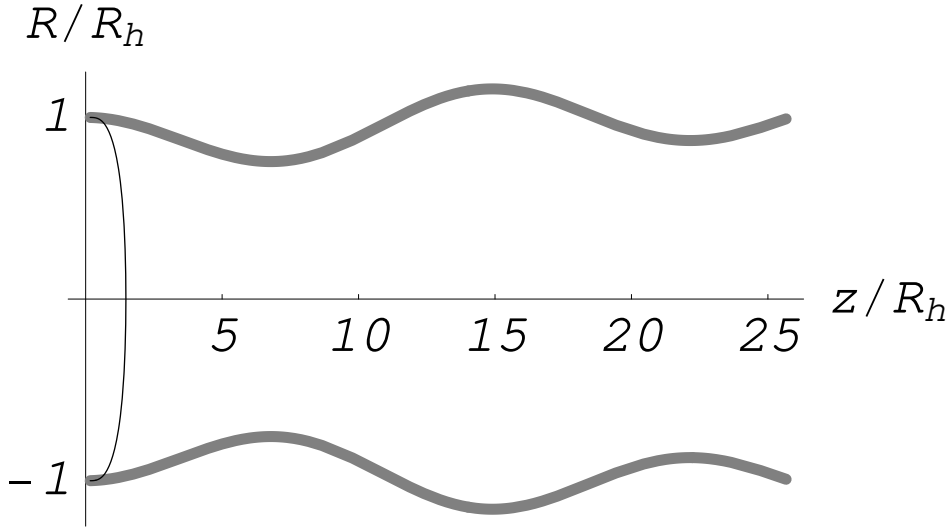,width=5in}}
\caption{ The horizon of the brane black hole from linearized gravity
 and the horizon of the perturbed black string. The perturbation shown
 is the unstable mode with the shortest wavelength.
\label{fig:profile}
}
\end{figure}

For a large black hole, the profile of the GL instability, given by
$f(z)$ in equation \eqref{eq:GL} goes as,
\begin{equation}
f(z) \sim J_2\left( k z \right) 
\label{eq:GLprofilebig}
\end{equation}
where we recall that the marginally unstable mode has eigenvalue $k_c
= 0.45/R_S$. We plot this profile in figure \ref{fig:profile}. Earlier
we estimated the shape of a large black hole using the linear theory,
finding that the coordinate extent, $\Delta z$, of the potential
isosurface we take to approximate the horizon into the bulk is $\Delta
z=2 R_S$. We also plot this in figure \ref{fig:profile}. We might take
the region of the isosurface extending a half or quarter of this
distance to approximate a uniform warped string. However, from figure
\ref{fig:profile}, it is then clear that only a tiny fraction of a
wavelength of the marginal mode could fit into this region. Hence we
find it very unlikely that any potential Gregory-Laflamme instability
could be localized in the uniform string region of the localized black
hole.

\subsection{Entropy Balance}

Even without finding the explicit black string instability, one could
have argued via entropy considerations that one would expect that the
black string is unstable.  We now apply this reasoning to the brane
black hole by comparing the entropy advantage of the localized black
hole to ``drip'' off a small black hole at the tip to the entropy of
the static solution. We find that from the perspective of entropy
considerations, the decay is parametrically marginal, so it cannot be
decided according to entropy considerations.

Consider a brane black hole that extends into the bulk out to some
value $z$ in $z$-coordinates which is close to $R_S$.  If the tip were
to drip off, then it would form a small black droplet of radius $L$,
and the brane black hole would shrink.  The black droplet will have
mass and entropy given approximately by
\begin{eqnarray}
\delta M_{drop} &\sim&
M_5^3 L^2
\label{dropm} \\
\delta S_{drop} &\sim& M_5^3 L^3
\label{drops}
\end{eqnarray}
The brane black hole will shrink by an amount given by energy
conservation.  The effective mass lost to the black droplet is
redshifted by the warp factor:
\begin{equation}
\delta M_{bbh} = - \delta M_{drop} \frac{L}{z}
\label{bbhm}
\end{equation}
The radius of the brane black hole then shifts by an amount $\delta
R_S = \delta M_{bbh}/M_4^2$, and the entropy shifts by
\begin{equation}
\delta S_{bbh} \approx M_5^3  L  \left[ \left( R_S - \delta R_S \right)^2 - R_S^2 \right] 
   \sim -M_5^3 L^3 \frac{R_S}{z}
\label{bbhs}
\end{equation}
For $z \approx R_S$, this is parametrically the same as the entropy of the
black droplet.  However, if the brane black hole sticks farther out
into the bulk, then the total entropy change $\delta
S_{drop} + \delta S_{bbh}$ grows and becomes positive, and formation of a black
droplet is favored.

Notice that for $z<R_S$, this analysis wouldn't apply since the black
holes that would be spit off in that case would be larger than the AdS
length scale, so the simple interpretation as pure five-dimensional
flat space black holes would certainly not apply. The analysis is only
valid up to the point where the entropy argument is indeterminate.

The formula for the entropy $S_{bbh}$ of the brane black hole deserves
a quick comment.  The black hole extends a proper distance 
into the bulk, so a naive estimate of the area would be
$S_{bbh} \sim R_S^2 L \log{R_S/L}$.  However, as the authors of
\cite{emparan:exact} discuss, the contribution to the area is
suppressed by the warp factor away from the brane, and almost all of
the area is near the brane itself.  A better estimate for the area is
\begin{equation}
d\mathcal{A} = \frac{4 \pi R_S^2 dz}{(z/L)^3}
\end{equation}
on each spatial slice of constant $z$.  Integrating $d\mathcal{A}$
from $z=L$ to $z=R_S$ accounts for the formula for $S_{bbh}$.

As there is no parametric argument for, or against this instability we
regard it as marginal. We expect that any modification of the bulk
physics might therefore effect this stability. For example, adding
charges to a uniform black string may render the GL
instability absent \cite{gubser-mitra}. Certainly extremally charged
localized black holes will be stable.

Another simple modification is to add extra compact dimensions.  In
this case, the area of the small black droplet can start to probe the
extra dimensions, whereas the large brane black hole will fill them
up.  Consider for example a brane black hole in $5+D$ dimensions where
$D$ of the dimensions are compact and of size $L$. The surface area of
a unit $3+D$-sphere is smaller than that of a $3$-sphere by the factor
\begin{equation}
\frac{\int d\Omega_{3+D}}{\int d\Omega_3} = \left( \frac{(2\pi)^{D/2}}{3\cdot
5\cdots (3+D-2)} \right)
\end{equation}
Since the black droplet in this scenario is roughly the same size as
the compact dimensions, it will interpolate between a $4+1$-dimensional
black hole and a $4+D+1$ dimensional black hole, and its entropy will pick up
some fraction of the above ratio.  By making $D$ large, it becomes
more likely that such the brane black hole is stable.

\section{Effective Theory Interpretation}

We now return to the issue mentioned in the introduction of why the
effective theory for linear fluctuations about a flat brane can be
given by pure 4-d gravity, but even for large radius black holes this
4-d gravity description can break down. We note that this is true
whether or not the black hole is unstable.  In either case, the
effective theory at the nonlinear level in the presence of a black
hole breaks down.

This might seem surprising since for the linear theory about the flat
brane we recover simple 4-d gravity for perturbations with wavelength
larger than $L$, the AdS length. Thus one might naively assume that
4-d gravity is the correct effective theory with a cut-off requiring
all curvature radii to be less than $L$. However, this is not the
case. Since the Kaluza-Klein spectrum is gapless, when we consider a
background containing a black holes with radius $R_S$ we should worry
about integrating out the modes with masses $m$ less than $\sim
1/R_S$. Usually theories that admit 4-d effective descriptions have a
mass gap in their spectrum and hence for large enough black holes
there will be no modes with such masses. Here however there is a
continuum of such light modes, down to zero mass. While the spectrum
of large mass modes, with masses $m >> 1/R_S$ will be unchanged due to
the presence of the black hole in the background, the spectrum of the
modes with masses approximately $m \sim 1/R_S$ will generically be
strongly affected.

  This modification of the effective theory is important and perhaps
more obvious when considering the \emph{quantum} radiation rate of the
5-d black hole.  From a purely 5-d perspective, there is only a single
graviton with $\mathcal{O}(1)$ degrees of freedom leading to a
relatively small radiation rate\cite{Emparan:2000rs}.  
However, if one were to calculate the decay rate using the 4d
theory with the original modes, one would find that the answer would
depend on an IR and a UV cut-off.  The resolution of this apparent
discrepancy is that the black hole has drastically changed 
the spectrum of KK modes.  The picture is
qualitatively the same as the usual calculation for Schwarzschild
black holes.  The phase space of graviton excitations in flat space
gets replaced in the presence of the black hole by spherical
harmonics, which are effectively quantized at the temperature of the
black hole, and oscillations in the radial direction.  In the case of
the localized black hole, the KK modes no longer have a continuous
spectrum, but effectively are quantized also at the temperature of the
black hole, leaving only $\mathcal{O}(1)$ accessible modes. So in
general, an effective theory with a KK spectrum continuous down to
zero can change in the presence of large geometric perturbations such
as a black hole. Note, however, that in the case of the brane black hole, modes 
with
mass lighter than the temperature have small overlap with the
higher-curvature region of the brane black hole.  Therefore only those
modes with mass of order the black hole temperature would be strongly
coupled.

Consider again our simple example of the warped uniform string from
section \ref{sec:uniform}. In the dual 4-d theory this does appear to
be exactly a 4-d Schwarzschild solution. However, as we have discussed
above, the Kaluza-Klein spectrum about this solution has tachyonic
modes, due to the GL instability. These are not present in the
spectrum of fluctuations in the absence of the black hole, and only
arise only through the non-linear interaction of the Kaluza-Klein
modes with the 4-d graviton. We see this non-linear interaction
explicitly in equation \eqref{eq:GLpert}, where the perturbation
governed by $\chi_{\mu\nu} f(z)$ is a Kaluza-Klein mode, and the
Lichnerowicz operator includes the non-linear coupling to the 4-d zero
mode, $g_{\mu\nu}(x)$, through $R^{(4)~\alpha~\beta}_{~~\mu~\nu}
\chi_{\alpha\beta}$ acting as a potential. Since the mass squared,
$-k^2$, may become arbitrarily small, when it becomes of order the
potential $\sim 1/R_S^2$, the 4-d mode behavior is strongly modified
from that of a usual 4-d field with the same mass.

Note that in addition to the 4d coupling of the graviton to the 
tower of KK modes, higher dimensional operators become more
important because of the large curvature components at the tip.
For both these reasons, conventional no-hair theorems do not apply.
Hence there can be two different types of black hole solutions - one
involving KK modes (the brane black hole we have been discussing)
and one without them (the black string).

\section*{Acknowledgments}

We would like to thank Nima Arkani-Hamed, Roberto Emparan, Hideaki
Kudoh, Lubos Motl, Robert Myers, and Veronica Sanz for very
interesting discussions. T.W. is grateful for the stimulating
environment and hospitality provided at the KITP workshop ''Scanning
New Horizons: GR Beyond 4 Dimensions''. The research of T.W. is
supported by NSF grant PHY-0244821.  L.R. and A.L.F. are supported by
NSF grants PHY-0201124 and PHY-0556111, and A.L.F. acknowledges an NSF
graduate research fellowship.

\end{document}